\documentstyle[12pt]{article}
\normalbaselines

\def\rf#1{(\ref{eq:#1})}
\def\lab#1{\label{eq:#1}}
\def\nonu{\nonumber}
\def\br{\begin{eqnarray}}
\def\er{\end{eqnarray}}
\def\be{\begin{equation}}
\def\ee{\end{equation}}

\def\foot#1{\footnotemark\footnotetext{#1}}
\def\lb{\lbrack}
\def\rb{\rbrack}
\def\llangle{\left\langle}
\def\rrangle{\right\rangle}
\def\blangle{\Bigl\langle}
\def\brangle{\Bigr\rangle}
\def\llb{\left\lbrack}
\def\rrb{\right\rbrack}

\def\lcurl{\left\{}
\def\rcurl{\right\}}
\def\({\left(}
\def\){\right)}
\def\v{\vert}                     
\def\bgv{\bigg\vert}              
\def\lskip{\vskip\baselineskip\vskip-\parskip\noindent}
\def\mskp{\par\vskip 0.3cm \par\noindent}
\def\sskp{\par\vskip 0.15cm \par\noindent}
\def\bc{\begin{center}}
\def\ec{\end{center}}

\relax

\def\Tr{\mathop{\rm Tr}}                  
\newcommand\partder[2]{{{\partial {#1}}\over{\partial {#2}}}}
\newcommand\funcder[2]{{{\delta {#1}}\over{\delta {#2}}}}
\newcommand\bil[2]{\left\langle {#1} \bigg\vert {#2} \right\rangle} 
\newcommand\me[2]{\left\langle {#1}\right|\left. {#2} \right\rangle} 

\newcommand\sbr[2]{\left\lbrack\,{#1}\, ,\,{#2}\,\right\rbrack} 
\newcommand\Sbr[2]{\Bigl\lbrack\,{#1}\, ,\,{#2}\,\Bigr\rbrack} 
\newcommand\Pbr[2]{\Bigl\{ \,{#1}\, ,\,{#2}\,\Bigr\}}  
\newcommand\pbbr[2]{\lcurl\,{#1}\, ,\,{#2}\,\rcurl}  

\def\a{\alpha}
\def\b{\beta}
\def\c{\chi}
\def\d{\delta}
\def\D{\Delta}
\def\eps{\epsilon}
\def\vareps{\varepsilon}
\def\g{\gamma}
\def\G{\Gamma}
\def\h{{1\over 2}}
\def\l{\lambda}
\def\L{\Lambda}
\def\m{\mu}
\def\n{\nu}
\def\ov{\over}

\def\O{\Omega}
\def\p{\phi}
\def\P{\Phi}
\def\pa{\partial}
\def\pr{\prime}
\def\s{\sigma}
\def\S{\Sigma}
\def\t{\tau}
\def\th{\theta}
\def\Th{\Theta}

\def\ti{\tilde}
\def\wti{\widetilde}

\def\cA{{\cal A}}

\def\cD{{\cal D}}

\def\cL{{\cal L}}
\def\cM{{\cal M}}
\def\cN{{\cal N}}
\def\cP{{\cal P}}
\def\cQ{{\cal Q}}
\def\cR{{\cal R}}
\def\cS{{\cal S}}
\def\cU{{\cal U}}
\def\cV{{\cal V}}
\def\cW{{\cal W}}
\def\cY{{\cal Y}}

\def\phanta{\phantom{Never Communism}}
\def\phantb{\phantom{Never Islamic Fundamentalism}}

\def\lie{{\cal G}}
\def\dlie{{\cal G}^{\ast}}

\def\ulie{{\cal U}\( {\cal G}\)}        

\font\upright=cmu10 scaled\magstep1
\def\Rmath{\vcenter{\hbox{\upright\rlap{I}\kern 1.7pt R}}}
\def\IR{\ifmmode\Rmath\else$\Rmath$\fi}
\def\one{\hbox{{1}\kern-.25em\hbox{l}}}
\def\0#1{\relax\ifmmode\mathaccent"7017{#1}%
        \else\accent23#1\relax\fi}

\def\Win1{{\bf W_{1+\infty}}}           
\def\nWinf{{\bf {\hat W}_\infty}}       
\def\DA{{\cal DOP} (S^1 )}                    
\def\eDA{{\widetilde {\cal DOP}}}     

\def\sto{\stackrel{\otimes}{,}}              
\def\sta{\, ,\,}
\def\xx{(\xi , x)}

\def\intres{\int dx\, {\rm Res}_\xi \; }
\def\Intres{\int dx\, {\rm Res} \; }

\def\PsDA{\Psi{\cal DO}}
\def\ePsDA{{\widetilde {\Psi{\cal DO}}}}
\def\dPsDA{\Psi{\cal DO}^{\ast}}

\def\eVolt{\Bigl( {\widetilde {\Psi{\cal DO}}} \Bigr)_{-}}

\def\eVOLT{\Bigl( {\widetilde {\Psi {\rm DO}}} \Bigr)_{-}}

\def\faa{Fa\'a di Bruno~}

\def\ot#1{{#1}\otimes \one + \one \otimes {#1}}
\def\dad{ad_{\ast}^{\ast}}        

\newcommand{\nit}{\noindent}
\newcommand{\ct}[1]{\cite{#1}}

\def\NPB#1#2#3{{\sl Nucl. Phys.} {\bf B#1} (#2) #3}
\def\NPBFS#1#2#3#4{{\sl Nucl. Phys.} {\bf B#2} [FS#1] (#3) #4}
\def\CMP#1#2#3{{\sl Comm. Math. Phys.} {\bf #1} (#2) #3}
\def\PRD#1#2#3{{\sl Phys. Rev.} {\bf D#1} (#2) #3}

\def\PLB#1#2#3{{\sl Phys. Lett.} {\bf #1B} (#2) #3}
\def\JMP#1#2#3{{\sl J. Math. Phys.} {\bf #1} (#2) #3}

\def\PTP#1#2#3{{\sl Prog. Theor. Phys.} {\bf #1} (#2) #3}

\def\AoP#1#2#3{{\sl Ann. of Phys.} {\bf #1} (#2) #3}

\def\PR#1#2#3{{\sl Phys. Reports} {\bf #1} (#2) #3}

\def\FAaIA#1#2#3{{\sl Functional Analysis and Its Application} {\bf #1}
(#2) #3}

\def\InvM#1#2#3{{\sl Invent. Math.} {\bf #1} (#2) #3}
\def\LMP#1#2#3{{\sl Letters in Math. Phys.} {\bf #1} (#2) #3}
\def\IJMPA#1#2#3{{\sl Int. J. Mod. Phys.} {\bf A#1} (#2) #3}

\def\TMP#1#2#3{{\sl Theor. Mat. Phys.} {\bf #1} (#2) #3}
\def\JPA#1#2#3{{\sl J. Physics} {\bf A#1} (#2) #3}

\def\MPLA#1#2#3{{\sl Mod. Phys. Lett.} {\bf A#1} (#2) #3}

\def\JETPL#1#2#3{{\sl  Sov. Phys. JETP Lett.} {\bf #1} (#2) #3}

\begin{document}
\vspace*{-1cm}
\noindent
\phantom{bla}  \hfill{\sl BGU-93 / 22 / October - PH} \\
\phantom{bla}  \hfill{\sl Revised ~December 93} \\
\phantom{bla}
\hfill{hep-th/9310113}
\\
\vskip .2in
\begin{center}
{\large {\bf STRING ~THEORY ~~AND ~~INTEGRABLE ~SYSTEMS\foot{To appear in
{\em ``Mathematical Physics Towards the XXIst
Century''}, R.N. Sen and A. Gersten eds. (Proc. of the Int. Conf. held at
Ben-Gurion Univ. of the Negev, Beer Sheva, March 14-19, 1993)}
}}
\end{center}
\begin{center}
Emil NISSIMOV${}^2$ ~and ~Svetlana PACHEVA\foot{On leave from the Institute
for Nuclear Research and Nuclear Energy of the Bulgarian Academy of
Sciences, Tsarigradsko Chausee 72, BG-1784 ~Sofia, Bulgaria; E-mail:
dstoyan@bgearn.bitnet; Fax: 359-2-755019 .}
\end{center}
\begin{center}
Department of Physics, Ben-Gurion University of the Negev, \\
P.O.Box 653, IL-84105 ~Beer Sheva, Israel \\
E-mail: emil@bguvms.bitnet; svetlana@chen.bgu.ac.il;
Fax: 972-57-281340
\end{center}

\lskip
{\large {\bf Abstract}} ~
This is mainly a brief review of some key achievements in a ``hot'' area of
theoretical and mathematical physics. The principal aim is to outline the
basic structures underlying {\em integrable} quantum field theory models with
{\em infinite-dimensional} symmetry groups which display a
radically new type of {\em quantum group} symmetries. Certain particular
aspects are elaborated upon with some detail: integrable systems of
Kadomtsev-Petviashvili type and their reductions appearing in matrix models
of strings; Hamiltonian approach to Lie-Poisson symmetries;
quantum field theory approach to two-dimensional relativistic
integrable models with dynamically broken conformal invariance.
All field-theoretic models in question are of primary relevance to
diverse branches of physics ranging from nonlinear hydrodynamics to string
theory of fundamental particle interactions at ultra-high energies.
\lskip
{\large {\bf 1. Introduction}}
\sskp
One of the prevailing views in modern theoretical physics is that
fundamental laws of Nature can be derived and understood in terms of
field-theoretic models in a lower dimensional space-time possessing
{\em infinite-dimensional symmetry groups} and, thus, as a rule
being {\em integrable}.

These models in their various facets and
disguises are encompassed in the extremely rich and rapidly developing
branch of {\em string theory} \ct{string}. It is widely believed that
string theory
is the most viable candidate for a unified theory of all fundamental
interactions at ultra-short distances which, in particular, will provide
a consistent reconciliation between General Relativity and Quantum
Mechanics - one of the major challenges of this century's Physics.

We have in mind two large classes of {\em integrable models} :
conformal field theory (CFT) \ct{BPZ84,CFT} and massive completely integrable
models \ct{integr,FT87,Zam's} in $D=2$ space-time dimensions.
Typical examples of CFT are the {\em rational} CFT's, of which the most
extensively studied are the Wess-Zumino-Novikov-Witten (WZNW) models for
various Lie groups $G$ and models, obtained from them by gauging of
different subgroups $H$ of $G$ \ct{RCFT}. Thoroughly studied models in the
second class are : Sine-Gordon, (nonabelian) massive Thirring models, Toda
models for various groups, the Korteveg-de Vries (KdV) and Kadomtsev-
Petviashvili (KP) integrable {\em soliton} evolution equations and
their hierarchies.
The feature of {\em integrability}, common for both classes of models,
stems from the infinite-dimensional Lie-algebraic structure they share :
in the first class this being the Noether symmetry algebra, and in the second,
the Hamiltonian structure.
The underlying infinite-dimensional symmetries, manifesting themselves
through the Virasoro (conformal) \cite{BPZ84} and affine Kac-Moody algebras
\cite{KM}, as well
as through various (infinite-dimensional) generalizations thereof - e.g.
the $W$-algebras \cite{W-alg}, play a crucial role in solving and
interpreting integrable models.

The interrelation between CFT and
completely integrable models became recently explicit through the appearance
of KdV and KP integrable hierarchies in the {\em matrix model}
description \cite{GMD} of (sub)critical strings ({\sl i.e.}, 2-dimensional
gravity interacting with conformal ``matter'' fields). Thus, it is precisely
the integrable field theories which provide the proper framework for
incorporation of the huge symmetries of string theory models.

The property of {\em integrability} is studied in two aspects :
{}~(1) classical and ~(2) quantum.
The principal questions, which one should answer, are the following :~~
(1a) classification; ~(1b) action-angle type of variables; ~(1c) exact
integration of the equations of motion --- in the classical case, and ~~
(2a) classification; ~(2b) exact scattering amplitudes;
{}~(2c) exact correlation functions of local fields off the
mass shell --- in the quantum case.

The concepts and tools to approach the above topics invoke, besides the
theory of infinite-dimensional Lie algebras and groups, another
outstanding branch of mathematical physics --
{\em symplectic geometry} or, equivalently, {\em Hamiltonian mechanics}.
The aim here is to uncover the {\em geometrical} foundations of the relevant
field theories. The use of geometric techniques offers powerful means
for unifying various physical theories and obtaining new insights.
It is just
in the realm of string theory and the theory of completely integrable
systems where the intertwining of Hamiltonian and Lie-group structures
in field-theoretic models attains new immensely important qualities.
Their quintessence is manifested in the development of the principal
methods to solve the quantization problems in integrable models :
{\em quantum inverse scattering method} \ct{QISM,Takht90},
representation theory of
infinite-dimensional Lie algebras \ct{KM}, {\em Quantum Groups}
(non-commutative and non-cocommutative Hopf algebras) \ct{QG,Takht90}.

The generic integrable models are massive field theories which in a
sense can be regarded as integrable perturbations of conformal
field theories \cite{Zam89}. Such models have the
advantage of being relativistic invariant and classifiable by the
conformal models of which they are perturbations. The latter
describe the renormalization group fixed points of
these massive integrable models.
Their most essential feature, explicitly exposing
the intimate connection to conformal models, is the existence of
{\em multi-Hamiltonian} structures, {\sl i.e.} the existence of at least a
second Hamiltonian structure which is compatible with the canonical
$R$-matrix Kirillov-Kostant structure \ct{STS83}. The
corresponding fundamental Poisson brackets (linear $R$-matrix brackets
and quadratic (Sklyanin) $R$-matrix brackets) naturally arise and are
exhaustively understood within the classical ``inverse scattering'' method
\cite{FT87}. Also, they can be deduced in the semiclassical limit from
the basic algebraic structures :
\sskp
(1) Yang-Baxter equation for the quantum version of the $R$-matrix;
\par\noindent
(2) Fundamental commutation relations for the quantum transfer matrix,
involving the quantum $R$-matrix as ``structure constants'',
\sskp
which enter the quantum ``inverse scattering''
method \cite{QISM} - the first systematic method for quantization of
completely integrable models.

In a related development, Drinfeld \cite{L-P-drin,QG} was the
first to realize the deeper algebraic and geometric nature built-in
into the theory of classical and quantum completely integrable models.
Namely, he showed that the algebraic structures (1) and (2), listed
above, constitute the basic structural relations of {\em
non-commutative and non-cocommutative Hopf algebras} which ultimately
received the name
{\em ``Quantum Groups''} and evolved into one of the ``hottest'' topics
in mathematics and theoretical physics. Furthermore, it was realized
that a quantum group is a {\em deformation} of a classical Lie group
much in the same way quantum mechanics is a deformation of classical
Hamiltonian (symplectic) mechanics \cite{Flato}.
Most importantly, in the
``semiclassical'' limit the basic quantum group algebraic structures
(1) and (2), formulated above, transliterate into a special distinguished
Hamiltonian structure on the classical Lie group $G$, called
{\em Lie-Poisson structure},
which is compatible with the group multiplication. This is precisely the
class of Hamiltonian structures given by the quadratic fundamental
$R$-matrix Poisson brackets mentioned above in the context of classical
completely integrable models.


The concept of  {\em quantum group symmetries} in integrable quantum
field and statistical mechanics' models lead in the recent years
to numerous fruitful and exciting developments in theoretical physics : from
generalization ({\sl i.e.}, {\em q-deformation})
of the fundamental notions of internal and space-time symmetries in
quantum field theory and spin-statistics connection to quantum magnetic chains
and critical dynamics \cite{QG-appl}. Furthermore, the concepts of
integrability and perturbations around exactly solvable theories
find their place in (close to) realistic models of elementary particles, e.g.
in quantum chromodynamics \ct{QCD}.

In what follows, few particular topics among those mentioned above are
discussed
in some detail.
\lskip
{\large {\bf 2. Matrix Models of Non-Perturbative Strings and Integrability }}
\mskp
{\bf 2.1 Conventional Perturbative String Theory}
\sskp
The standard geometric formulation of {\em perturbative} string theory
\ct{Pol81} provides the following prescription for calculating physical
observables (fermionic degrees of freedom are discarded for simplicity) :
to construct scattering amplitudes one considers functional
integrals over (Euclidean) string world-sheets $\S_{A,G}$ -- smooth Riemann
surfaces embedded in $D$-dimensional (Euclidean) space-time $ R^D$ of genus
$G$ and area $A$ :
\br
Z_{string} = \sum_{G=0}^{\infty} g^{2G} \int dA \, e^{-\L A} Z_{A,G}
\phantb    \lab{1} \\
Z_{A,G} = {\Large\int} \llb \cD h \rrb \cD X \, \exp \lcurl
- S_{string} \lb X,h\rb \rcurl \,
\prod_{i=1}^n \int d^2 \s_i \sqrt{h} \, V \( X,h; k_i \)   \lab{2}
\er
\be
S_{string} =\h \int d^2 \s \llb \sqrt{h} \(
h^{ab} \pa_a X^\m \pa_b X^\n G_{\m\n} (X) + \P (X) R^{(2)} (h) \) +
\eps^{ab} \pa_a X^\m \pa_b X^\n B_{\m\n} (X) \rrb  \lab{3}
\ee
Here the following notations are used : $g$ denotes the string ``coupling''
constant, $\L$ is the ``cosmological'' constant, the string action
$S_{string}$ represents a typical $D=2$ conformally invariant field theory
model describing $D=2$ gravity (given by the world-sheet metric $h_{ab}(\s )$
) coupled to world-sheet ``matter'' fields $X^\m (\s )$ (describing the
embedding of $\S_{A,G}$ in $R^D$ ). The functionals $ G_{\m\n}, \P ,B_{\m\n}$
represent the space-time dilaton-gravity multiplet. String interactions are
given in \rf{1},\rf{2} by geometrical splitting and joining of individual
string world-sheets (thus creating handles on the total world-sheet), whereas
asymptotic in-coming and out-going string states are given by vertex operators
$V (X,h;k)$ ($k$ indicating the momentum of in/out-state).

Henceforth, for simplicity, we shall suppress the vertex operators $V$'s in
\rf{2}, {\sl i.e.}, we shall concentrate on the string partition function.

The most difficult part in calculating \rf{2} is the accurate treatment of the
functional measure $\llb \cD h \rrb$ over the space of all {\em
gauge-inequivalent} classes of metrics $h_{ab}$ on $\S_{A,G}$ w.r.t.
reparametrization and Weyl conformal invariance. In case of conformal gauge
$ h_{ab} = e^{\p} {\hat h}_{ab}(\t )$ where $\p$ is the Weyl conformal
factor and ${\hat h}_{ab}(\t )$ is a reference metric with constant curvature
$R^{(2)}\( {\hat h}_{ab}\)$ and which depends in general on the moduli
$\{ \t \}$ of the corresponding Riemann surface $\S_{A,G}$ , the standard
Faddeev-Popov gauge-fixing procedure yields \ct{Pol81,OAlv83} :
$\llb \cD h \rrb = \d \( h_{ab} - e^{\p} {\hat h}_{ab}(\t ) \) ~\D_{\P\Pi}
{}~\cD \p ~\Bigl( d\t \Bigr) $ ,
with the Faddeev-Popov determinant $\D_{\P\Pi}$ giving rise to the
well-known conformal anomaly.

An important result about the entropy of (random) surfaces with fixed area
$A$ , first obtained by Zamolodchikov \ct{Zam82} in the
semi-classical approximation and subsequently strengthened in \ct{KPZ+},
states that for large $A$ :
\be
Z_{A,G} ~{\simeq}_{A \rightarrow \infty} ~{\rm const}_G ~
e^{\L_c A} ~A^{-\c \( 1-\g_0 /2 \) -1}  \lab{5}
\ee
where $\L_c$ denotes a ``critical'' value of the ``cosmological''
constant $\L$ ,
$\c = 2 (1-G)$ is the Euler characteristics of the surfaces, and $\g_0$
denotes a critical exponent depending on the world-sheet ``matter'' fields.
Relation \rf{5} implies for the string partition function \rf{1} :
\be
Z_{string} \simeq \( \L - \L_c \)^{2-\g_0} \sum_{G=0}^{\infty} \Bigl( const
\Bigr)_G
\( {{g^2} \over { \( \L - \L_c \)^{2-\g_0}}} \)^G    \lab{6}
\ee
which shows that one can obtain complete nonperturbative result for
$Z_{string}$ by taking the {\em double scaling} limit :
\be
\L \longrightarrow \L_c \quad , \quad  g^2 \longrightarrow 0 \qquad
{\rm such ~that} \qquad g^2_{ren} \equiv
\( {{g^2} \over { \( \L - \L_c \)^{2-\g_0}}} \) = {\rm fixed} \lab{7}
\ee
\mskp
{\bf 2.2 Lattice Regularization of String Theory; Matrix Model Formulation}
\sskp
In a series of pioneering papers \ct{triang} it was found that statistical
mechanical models of random matrices (``matrix models'' for short)
provide an adequate apparatus for nonperturbative
description of lattice-regularized string theory based on the method of
{\em random triangulation} (and, more generally, random polygonization)
of the (Euclidean) string world-sheet. A decisive breakthrough occurred
further in refs.\ct{GMD} which proposed ways for
correct implementation of the continuum limit as
double scaling limit \rf{7} allowing for exact solutions in string theory.

Let us note that, whereas matrix model formulation of random surfaces
is adequate for solving integrable lattice
models of planar statistical mechanics, its application to genuine string
theory is limited so far to the case of $D \leq 2$ dimensional embedding space.
Nonetheless, the exact solvability of matrix models provides an important
testing ground and qualitative hints for the nonpertubative string theory
solution in the realistic cases (for extensive reviews, see \ct{MMrev}).

Since our primary goal here is to elucidate the emergence of integrability
structures in the matrix models of string theory, we shall
consider for illustrative purpose the simplest one-matrix model
whose partition function is given by :
\be
Z = \int d^{N^2}M \, e^{-N \, V(M)} \qquad , \qquad
V(M) = \sum_{k \geq 0} t_k \( \frac{N}{\b} \)^{k/2 -1} \Tr M^k  \lab{8}
\ee
with $M = \| M_{ij} \|$ being a $N \times N$ hermitian matrix. In ordinary
perturbation theory defined in terms of a free ``propagator''
$\llangle M_{ij} M_{kl} \rrangle_{(0)} \sim N^{-1} \d_{ik} \d_{jl}$ and $k$-leg
vertices with weights $ t_k N \( \frac{N}{\b} \)^{k/2 -1}$ , each diagram
$\G$ gives contribution of the form :
\be
\prod_{k \geq 3} \( t_k N \( \frac{N}{\b} \)^{k/2 -1} \)^{V_k (\G )} \ldotp
N^{- P(\G )} \ldotp N^{L(\G )} =
\cW_\G \llb\lcurl t \rcurl\rrb \, N^{\c (\G )} \,
\( \frac{\b}{N} \)^{-L(\G )}    \lab{9}
\ee
On the l.h.s. of \rf{9} $\, V_k (\G ) \, , P(\G ) \, , L(\G )$ denote number of
$k$-leg vertices, propagators (links) and closed loops (faces) of $\G$ ,
whereas on the r.h.s. $\, \c (\G ) =
L(\G ) - P(\G ) + \sum_{k \geq 3} V_k (\G )\, $ denotes the Euler
characteristics of the two-dimensional polygonized surface spanned by $\G$ ,
and ${\cal W}_{\G} \llb\lcurl t \rcurl\rrb $ indicates the product of
the vertex weights. Clearly, $L(\G ) \equiv A(\G )$ can be understood as area
of $\G $ . Thus, the partition function \rf{8} can be written as :
\be
Z = \exp \lcurl \sum_{{\rm conn. ~surfaces} ~\G} N^{\c (\G )}
e^{ -\( \ln \b/N \) A (\G ) + \ln {\cal W}_{\G} \llb \{ t\} \rrb} \rcurl
\lab{10}
\ee
{\sl i.e.}, the {\em free energy} $\ln Z$ of the matrix model \rf{10}
represents
the discretized regularized partition function of random surfaces (more
precisely, ``pure'' $D=2$ gravity with action $A(\G )$ interacting
with ``matter'' with action $\ln \cW_\G$ ) upon making the
following identifications (comparing with \rf{1}-\rf{2}, \rf{5}) :
$\, 1/N \simeq g$ (``bare'' string coupling constant), $\, N/\b \simeq
e^{-\( \L - \L_c \)} $ ($\L$ - the ``cosmological'' constant) and the
double scaling limit \rf{7} takes the form of a special continuum limit :
\be
\frac{N}{\b} \longrightarrow 1 \quad , \quad N \longrightarrow \infty
\quad , \quad {\rm such ~that} \quad g^2_{string} \equiv
\llb \b^2 \( \frac{\b}{N} - 1\)^{2-\g_0} \rrb^{-1} = {\rm fixed} \lab{11}
\ee

Explicit solution for the partition function \rf{8} is derived
by the method of orthogonal polynomials \ct{ortho}. Diagonalizing the
hermitian matrix $ M = U {\rm diag}\( \l_1 ,\ldots ,\l_N \) U^{-1}$ and
integrating over angle variables in $U$ , one gets (rescaling
$M \longrightarrow \( \b/N \)^{\h} M$ ) :
\be
Z =  \int \prod_{i=1}^N  d\l_i \, \D (\l ) \exp \lcurl - \b \sum_{i=1}^N
V \( \l_i \) \rcurl \D (\l ) \quad ; \; V(\l_i ) = \sum_{k \geq 0}
t_k \l_i^k \; , \; \D (\l ) = \prod_{i < j} \( \l_i - \l_j \)  \lab{12}
\ee
Introducing a complete set of orthogonal polynomials $P_n (\l )= \l^n +$
lower order terms :
\be
\int d\l P_n (\l ) e^{- \b V (\l )} P_m (\l ) = h_n \d_{nm}     \lab{13}
\ee
and re-expressing the van der Mond determinant as $\D (\l ) =
det \| P_i (\l_j )\|$ , the integrals in \rf{12} factorize and yield upon using
\rf{13} :
\be
Z \llb \{ t \} \rrb = \prod_{n=0}^N h_n \( \{ t \} \)    \lab{14a}
\ee
where the dependence on the parameters of the random matrix potential is
indicated. Explicit solutions for $ h_n \( \{ t \} \)$ can be found through
solving the flow equations w.r.t. $ t_k$ which correspond to integrable
lattice hierarchies, as briefly discussed in the next subsection.
\mskp
{\bf 2.3 Differential Integrable Hierarchies from Matrix Models}
\sskp
The appearance of integrable hierarchies in the continuum (double scaling)
limit \rf{11} are extensively discussed in numerous papers (see \ct{MMrev} and
refs. therein).
It turns out, however, that flow equations inherent to integrable hierarchies
can be extracted directly from discrete matrix models even before taking the
continuum limit \ct{Martinec,BX1}, which reveals their close connection
with topological field theories \ct{topolog}.

The above result is achieved most easily by employing again the method of
orthogonal polynomials \rf{13}. Namely, on the Hilbert space spanned by
$\lcurl P_n (\l ) \rcurl_{n \geq 0}\, $ one introduces two conjugate
operators $ \cQ , \cP$ with matrix elements defined as :
\br
h_n \cQ_{mn} = \int d\l P_n (\l ) e^{- \b V (\l )} \l P_m (\l ) \lab{14} \\
h_n \cP_{mn} = \int d\l P_n (\l ) e^{- \b V (\l )} \frac{d}{d\l} P_m (\l )
\lab{15}
\er
{}From \rf{14}, \rf{15} one easily gets the matrix model ``string'' equation
(the second one below) :
\be
\cP = \b \( V^{\pr} (\cQ ) \)_{(-)} \quad \longrightarrow \quad
\llb \b \( V^{\pr} (\cQ ) \)_{(-)} , \,\cQ \rrb = \one    \lab{16}
\ee
where the subscript $(-)$ denotes taking strictly lower-diagonal part of the
corresponding matrix. The ``string'' equation yields recurrsion relations for
the matrix elements \rf{14} of $\cQ$ .

It is straightforward to deduce from \rf{14} and \rf{13} the following
flow equations :
\br
\partder{P_n (\l )}{t_r} = \cQ^r_{(-)\; nm} P_m (\l ) \qquad , \qquad
\cQ_{nm} P_m (\l ) = \l P_n (\l )   \lab{17}   \\
\partder{\cQ}{t_r} = \sbr{\cQ^r_{(-)}}{\cQ} \phanta  \lab{18}
\er
Eq.\rf{18} is the integrability condition for eqs.\rf{17} and it is compatible
with the ``string'' equation \rf{16}. One can easily identify \rf{18} with the
Lax form of the flow equations of the integrable Toda lattice hierarchy by
inserting into \rf{18} the explicit form of the $\cQ$ matrix elements :
\be
\cQ_{n,n+1} =1 \quad ,\quad \cQ_{n,n} \equiv \cS_{n-1}  \quad , \quad
\cQ_{n+1,n} = \frac{h_{n+1}}{h_n} \equiv \cR_n
\equiv e^{\p_n - \p_{n-1}}\lab{19}
\ee
the rest being zero as a consequence of the recurrence relations for orthogonal
polynomials. Eqs.\rf{18} are now Hamiltonian, and the lowest one (with $r=1$)
is generated by the well-known Toda lattice Hamiltonian :
\be
H_{Toda} = \h \sum_n \cS_n^2 + \sum_n \( e^{ \p_{n+1} - \p_n} -1 \) \qquad ,
\qquad  \pbbr{\cS_n}{\p_m} = \d_{nm}  \lab{20}
\ee
Following \ct{BX1}, it is possible to replace the discrete lattice integrable
hierarchy \rf{18} by a differential hierarchy at each fixed lattice site $n$
where the continuum variable is $ x = t_1$ . Indeed,
from the first ($r=1$) flow eqs. \rf{17} and \rf{18} yielding
\be
\partder{P_{n+1}}{t_1} = \cR_n P_n \quad , \quad \partder{\cR_n}{t_1} =
\cR_n \( \cS_{n-1} - \cS_n \) \quad , \quad \partder{\cS_n}{t_1} =
\cR_n - \cR_{n+1}    \lab{21}
\ee
one obtains (upto gauge transformation  $\, P_n \longrightarrow \psi_n =
\exp \lcurl \int dt_1^{\pr} \, \cS_{n-1} \rcurl P_n = h_n^{-1} P_n $ ,
and similarly $ \cQ_{nm} \longrightarrow {\hat \cQ}_{nm} =
h_n^{-1} \cQ_{nm} h_m  $ ) :
\be
\l \psi_n = {\hat \cQ}_{nm} \psi_m =
h_n^{-1} \( P_{n+1} + \cS_{n-1} P_n + \cR_{n-1} P_{n-1} \)
=  \llb \pa + \cR_n \( \pa - \cS_n \)^{-1}\rrb \psi_n        \lab{22}
\ee
where $\pa \equiv \partder{}{t_1}$ . Once again using \rf{21}, one can rewrite
the discrete evolution eqs.\rf{17},\rf{18} in a differential Lax form for a
fixed lattice site $n$ :
\be
\partder{\psi}{t_r} = \( L^r \)_{+} \psi   \qquad , \qquad
\partder{L}{t_r} =  \sbr{\( L^r \)_{+}}{L}              \lab{23}
\ee
where $r \geq 2 \; ,\; \psi \equiv \psi_n (t_1 ) $ and the subscript $+$
indicates taking the purely differential part of the corresponding
pseudo-differential Lax operator (cf. last equality in \rf{22}) :
\be
L = \pa + A \( \pa - B \)^{-1} \qquad , \qquad  A \equiv \cR_n (t_1 )
\quad , \quad  B \equiv \cS_n (t_1 )      \lab{24}
\ee
Eqs. \rf{23}, \rf{24} are immediately recognized as the well-known
2-boson reduction of KP integrable hierarchy.

Using generalization of the method of orthogonal polynomials, it is possible
to derive flow equations for integrable hierarchies also in the general case
of multi-matrix models (describing random surfaces interacting with $q$
different types of ``matter'') :
\be
Z = {\large\int} \prod_{i=1}^{q} d^{N^2}M_i \;
\exp \lcurl - \sum_{i=1}^q \( \Tr M_i M_{i+1} + \sum_{k \geq 0}
t_{i,k} \Tr M_i^k \) \rcurl          \lab{24a}
\ee
without passing to the continuum limit \ct{BX1}.
The appropriate generalization of \rf{24} now reads :
\be
L_q = \pa + \sum_{l=1}^q A_l \( \pa - B_l \)^{-1} \( \pa - B_{l+1} \)^{-1}
\ldots \( \pa - B_q \)^{-1}     \lab{25}
\ee
which is a $2q$-boson reduction of KP integrable hierarchy
(see subsection 3.3 for more details).

Finally, returning to the string partition function \rf{14a}, one can show
that $Z \llb \{ t \} \rrb  = \t_N (t) $ -- the Toda lattice $\t$-function
\ct{Martinec} subject to the so called Virasoro constraints :
\br
\cL_s Z \llb \{ t \} \rrb  = 0 \quad , \quad s\geq -1 \qquad ; \qquad
\sbr{\cL_r}{\cL_s} = (r-s) \cL_{r+s}  \lab{26}  \\
\cL_{s \geq 0} = \sum k t_k \partder{}{t_{k+s}} + \sum_{k=0}^s
\partder{}{t_k} \partder{}{t_{s-k}}    \quad , \quad
\cL_{s<0} = \sum k t_k \partder{}{t_{k-s}} - \sum_{k=0}^{|s| -1}
\partder{}{t_{-k}} \partder{}{t_{s+k}}             \nonu
\er
which are equivalent to the constraints on the pertinent integrable hierarchies
imposed by the ``string'' equation \rf{16} and are in fact Ward identities
due to the symmetry $ \d_s M = \vareps_s M^{s+1} \; , \; s \geq -1$ of \rf{8}.
Similar relations hold for multi-matrix models. Moreover, in the continuum
(double scaling) limit $Z \llb \{ t \} \rrb $ can be identified with
$\t$-functions of reduced KP hierarchies subject to the so-called
$W$-constraints (generalizations of \rf{26} which span $W$-algebras;
cf. \rf{Winf}-\rf{omega} below).
\lskip
{\large {\bf 3. Integrable Systems in Classical Physics : ~Geometric
Formulation}}
\sskp
Now we pass to a brief review of some principal structures and properties
of integrable systems in classical and quantum mechanics (or field theory).
We first recall the notion of integrability.
\mskp
{\sl Complete integrability:}
Consider a Hamiltonian system with $n$ degrees of freedom possessing
standard Hamiltonian structure with a Hamiltonian $\, H(p,q)$ and Poisson
bracket $\{ \cdot,\cdot \}$.
A Hamiltonian system is called {\em completely} (or Liouville) {\em integrable}
if it has $n$ independent conserved quantities (integrals of motion) $I_k
(p,q)$
, $k=1, \ldots ,n$, which are in involution: $ \{ I_i , I_j \} = 0 $.
For such a system we can find the {\em action-angle} canonical variables and
write explicitly the general solution to the equations of motion.
\mskp
{\sl Lax formulation:}
For infinite-dimensional (field theory) integrable Hamiltonian systems ,
there exists the convenient Lax (or ``zero-curvature") formulation \ct{FT87}.
In the Lax formulation, the phase space of the Hamiltonian system is
parametrized by elements $L$ taking values in some Lie algebra $\lie$ and
the dynamical equations of motion can be written in
terms of a Lax pair $L$, $P$, the latter similarly taking values in $\lie$, as
the Lax-type equation
\be
{ d L \ov dt} = \lb L\, , \, P \rb  \lab{laxeq}
\ee
The Lax formulation leads straightforwardly to the construction of the
involutive integrals
of motion. Namely, for any Ad-invariant function $I$ on $\lie$, $I \( L\) $
is a constant of motion.
In fact, it can be shown that any completely integrable Hamiltonian system
admits a Lax representation (at least locally) \ct{BV90}.
\mskp
{\bf 3.1 Adler-Kostant-Symes/Reyman-Semenov-Tyan-Shansky Scheme}
\sskp
A very wide class of integrable models can be constructed through the
application of the AKS-RS approach \ct{AKS,STS83} having roots in the group
coadjoint orbit method \ct{KK}.

Let $G$ denote a Lie group and $\lie$ be its Lie algebra.
$G$ acts on $\lie$ by the adjoint action:
$ Ad(g) \, X = g X g^{-1}$, with $ g \in G$ and $X \in \lie$.
Let $\dlie$ be the dual space of $\lie$
relative to a non-degenerate bilinear form $\langle \cdot \v
\cdot \rangle$ on $\dlie \times \lie$.
The corresponding coadjoint action of $G$ on $\dlie$ is obtained from
the duality of $\langle \cdot \v \cdot \rangle$: $ \langle Ad^{\ast}(g) U
\v X \rangle = \langle U \v Ad (g^{-1}) X \rangle $.
We will denote the infinitesimal versions of adjoint and coadjoint
transformations (for $ g = \exp Y$) by $ad (Y)X = \sbr{Y}{X}$ and
$\me{ad^{\ast} (Y) U}{X} = -\me{U}{\sbr{Y}{X}}$ , respectively. In
particular, when $\lie$ is endowed with an $ad$-invariant bilinear
(Killing) form $ (\cdot ,\cdot )$ allowing to identify $\dlie$ with
$\lie$ , we have $ ad^{\ast} (Y) U = \sbr{Y}{U}$ .

There exists a natural Poisson structure on the space
$C^{\infty} \(\dlie, \IR \)$ of smooth, real-valued functions on
$\dlie$ (sometimes called Kirillov-Kostant (KK) bracket), given by :
\be
\{ F \, , \, H \} (U) =
- \bil{U}{\Sbr{\nabla F (U)}{\nabla H (U)}}     \label{eq:LBKKbra}
\ee
where $F, H \in C^{\infty} \(\dlie, \IR \)$ , the gradient
$\nabla F\,: \, \dlie \longrightarrow \lie$ is defined by the
standard formula ${d \ov dt} F \left( U + t V \right)\vert_{t=0} =
\llangle V \, \vert \, \nabla F (U) \rrangle$
and $\lb \cdot \, , \, \cdot \rb $ is the standard Lie commutator on $\lie$.
On each orbit of $G$ in $\dlie$ the Poisson bracket \rf{LBKKbra} gives rise to
a
non-degenerate symplectic structure.
Moreover, for any Hamiltonian function $H$ on such an orbit we have a
Hamiltonian equation of motion $ d U / dt = ad^{\ast} (\nabla H (U)) U
\; (\, = \sbr{\nabla H (U)}{U}\,$ when $\lie$ has Killing form) .

Ref.\ct{STS83} introduced the $R$-operator (generalized
$R$-matrix) as a linear map from a Lie algebra $\lie$
to itself such that the bracket:
\be
\lb X , Y \rb_{R} \equiv \h \lb R X , Y \rb + \h \lb X, R Y \rb
\lab{R-lie}
\ee
defines a second Lie-bracket structure on $\lie\,$, or equivalently,
defines a second Lie algebra $\lie_R$ isomorphic to $\lie$ as a vector
space.
The Jacobi identity for the $R$-commutator \rf{R-lie} implies that the
modified {\em Yang-Baxter equation} (YBE) for the $R$-matrix must hold
(for arbitrary $ X_{1,2,3} \in \lie$ ) :
\be
\sum_{{\rm cyclic} (1,2,3)} \biggl\lb X_1 \, ,\, \Sbr{R X_2}{R X_3}
- R \( \Sbr{R X_2}{X_3} + \Sbr{X_2}{R X_3} \) \biggr\rb = 0  \lab{mYBE}
\ee
A sufficient condition (occurring in all interesting cases) for the fulfilment
of \rf{mYBE} is :
\be
\Sbr{R X}{R Y} - R \( \Sbr{R X}{Y} + \Sbr{X}{R Y} \) = -\a \Sbr{X}{Y}
\lab{mYBE1}
\ee
($\a$ being arbitrary constant), which usually is written
in terms of the ``ordinary" $r$-matrix $\, r \in \lie \otimes \lie$ canonically
isomorphic (upto a factor) to $ \, R \in \lie \otimes \dlie$ via the Killing
form \foot{$ r = r_{ij} T^i \otimes T^j$ and $ \h R X = T^i r_{ij} \( T^j,X\)$
, where $\lcurl T^i \rcurl$ denotes a basis in $\lie$ .} :
\br
\Sbr{\stackrel{(12)}{r}}{\stackrel{(13)}{r}} +
\Sbr{\stackrel{(13)}{r}}{\stackrel{(23)}{r}} +
\Sbr{\stackrel{(12)}{r}}{\stackrel{(23)}{r}} =
ad{\rm -invariant}   \lab{CYBE} \\
\stackrel{(12)}{r} \equiv r_{ij} T^i \otimes T^j \otimes \one \in
\ulie \otimes \ulie \otimes \ulie \quad {\rm etc.}  \nonu
\er
where $\ulie$ is the universal enveloping algebra. Classification of solutions
of \rf{CYBE} (depending on a spectral parameter) for simple $\lie\,$ is given
in refs.\ct{BelDrin82}.

With the help of \rf{R-lie} one can furthermore introduce on
$\dlie_R \simeq \dlie \,$ a new KK-type Poisson bracket $\{ \cdot, \cdot\}_{R}$
called $R$-bracket by substituting the usual Lie commutator
$ \lb \cdot , \cdot\rb $ for the $R$-Lie commutator
$ \lb \cdot , \cdot\rb_{R} $ \rf{R-lie} in \rf{LBKKbra}:
\be
\{ F \, , \, H \}_R (U) =
- \bil{U}{{\Sbr{\nabla F (U)}{\nabla H (U)}}_R }    \lab{Rbra}
\ee
A function $H$ on $\dlie$ is called $Ad^{\ast}$-invariant
(Casimir) if $\; H\lb Ad^{\ast}(g) U\rb = H\lb U\rb\;$ or,
infinitesimally, $ ad^{\ast} (\nabla H (U)) (U) =0$ for each $U \in \dlie$.
Then one can show \ct{STS83} that: \\
\nit
(1) the $ad^{\ast}$-invariant functions are in involution
with respect to both brackets \rf{LBKKbra} and \rf{Rbra};\\
\nit
(2) the Hamiltonian equation on $\dlie \simeq \dlie_R\,$ takes the following
(generalized Lax) form:
\be
d U /dt = \h ad^{\ast} \Bigl(R \(\nabla H (U)\)\Bigr) U \quad \biggl(
= \Sbr{\h R \( \nabla H (U)\)}{U} \quad {\rm for}\; \lie \;{\rm with ~
Killing ~form} \biggr)     \lab{rhameqs}
\ee
Eq.\rf{rhameqs} can be obtained from
variational principle with the following geometric action:
\br
\cW \lb U\rb = - \int \bil{U}{\cY_R (U)} -\int dt \, H \lb U\rb
\lab{R-action} \\
dU = ad^{\ast}_R \( \cY_R (U) \) U \quad \longrightarrow  \quad
d \cY_R = \h \llb \cY_R \, ,\, \cY_R \rrb_R    \lab{MC}
\er
where the integrals in \rf{R-action} are along arbitrary smooth curve
on the phase space $\dlie$ ; $ H \lb U\rb $ is any Casimir on $\dlie$ ;
$ \cY_R (U)$ is Maurer-Cartan one-form on
$\lie_R$ and function of $U \in \dlie$ determined by the first eq.\rf{MC},
with the $R$-coadjoint action :
\br
ad^{\ast}_R (X) U = \h \Bigl( ad^{\ast}(RX) U
+ R^{\ast}\bigl( ad^{\ast}(X) U \bigr) \Bigr)  \phanta \nonu  \\
\biggl( \; =
\h \Sbr{RX}{U} - \h R \Bigl(\Sbr{X}{U} \Bigr) \quad {\rm for}\; \lie \;{\rm
with ~
Killing ~form} \biggr)      \lab{R-coadj}
\er

Hence the above AKS-RS technique leads to a direct construction of
completely integrable systems where the set of independent Casimir functions on
$\dlie$ is precisely the complete set of integrals of motion in involution.
The basic realization of this technique arises when the Lie algebra $\lie$
decomposes as a vector space into two subalgebras $\lie_{+}$ and
$\lie_{-}$, {\sl i.e.} $\lie = \lie_{+} \oplus \lie_{-}$. Let $P_{\pm}$ be the
corresponding projections on $\lie_{\pm}$. Then $R= P_{+} - P_{-}$
satisfies the modified YBE \rf{mYBE} and provides a specific realization
for the above scheme. In this case eqs.\rf{R-lie}, \rf{Rbra}, \rf{rhameqs} and
\rf{R-coadj} take the following simple form:
\br
{\Sbr{X}{Y}}_{R} = \Sbr{X_{+}}{Y_{+}} - \Sbr{X_{-}}{Y_{-}} \quad , \quad
\Bigl( ad_R^{\ast}(X) U \Bigr)_{\pm} = \mp {\Sbr{X_\mp}{U_\pm}}_\pm
\lab{R-lie1} \\
\pbbr{\bil{U_{\mp}}{X_{\pm}}}{U_{\mp}} = \pm {\Sbr{X_{\pm}}{U_{\mp}}}_{\mp}
\quad , \quad
{{dU}\ov {dt}} + \llb \( \funcder{H}{U} \)_{+},\, U \rrb = 0  \lab{Rbra1}
\er
where $X_{\pm}= P_{\pm} X \in \lie_{\pm} \; ,\; U_{\pm} = P^{\ast}_{\mp} U
\in \( \lie_{\mp} \)^{\ast} \; , \; {\sbr{X}{U}}_{\pm} = P^{\ast}_{\mp}
\Bigl( \sbr{X}{U} \Bigr) \in \( \lie_{\mp} \)^{\ast} $ .
\mskp
{\bf 3.2 Algebra of Pseudo-Differential Operators and Integrable Hierarchies of
Kad\-om\-tsev-Petvia\-shvili type}
\sskp
Here we will illustrate the AKS-RS construction on $\lie = \PsDA\,$ - the
algebra of pseudo-differential operators on a circle. Recall
that an arbitrary
pseudo-differential operator $X (x, D_x ) = \sum_{k \geq -\infty}
X_k (x) D_x^k \,$ is conveniently represented by its symbol
\ct{treves} - a Laurent series in the variable $\xi$ :
$X \xx = \sum_{k \geq - \infty} X_k (x) \xi^k $ ,
and the operator multiplication corresponds to the following symbol
multiplication:
\be
X \xx \circ Y \xx = \sum_{N \geq 0} {1 \ov {N !}}
{{\pa^N X} \ov {\pa \xi^N}} {{\pa^N Y} \ov {\pa x^N}}      \lab{product}
\ee
which determines a Lie algebra structure given by :
$\lb X , Y \rb \equiv \( X \circ Y - Y \circ X \)$.

On $\PsDA\,$ one introduces an invariant, non-degenerate bilinear form:
\be
\blangle L \v X \brangle \equiv {\Tr}_A \( L X \) =
\intres \Bigl( \, L\xx \circ X\xx \, \Bigr)         \lab{adler}
\ee
which allows identification of the dual space $\dPsDA$ with $\PsDA$ .

There exist three natural decompositions of $\lie = \PsDA$ into a linear sum of
two subalgebras $\lie = \lie^{\ell}_{+} \oplus \lie^{\ell}_{-} $
labelled by the index $\ell$ taking values $\ell = 0,1,2$:
\be
\lie^{\ell}_{+} = \lcurl X_{+} \equiv X_{\geq \ell}
= \sum_{i=\ell}^{\infty} X_i (x) D^i\rcurl
\qquad;\qquad
\lie^{\ell}_{-} = \lcurl X_{-} \equiv X_{< \ell}
= \sum_{i=-\ell+1 }^{\infty} X_{-i}(x) D^{-i} \rcurl \, ,
\lab{subalg}
\ee
Correspondingly the dual spaces to subalgebras $\lie^{\ell}_{\pm}$ are given
by:
\be
{\lie^{\ell}_{+}}^{\ast} = \lcurl L_{-} \equiv L_{< -\ell}
= \sum_{i=\ell+1}^{\infty}
D^{-i} \circ u_{-i}(x)\rcurl \;\;\; ;\;\;\;
{\lie^{\ell}_{-}}^{\ast} = \lcurl L_{+} \equiv L_{\geq -\ell} =
\sum_{i=-\ell}^{\infty} D^{i} \circ u_{i}(x) \rcurl \; .
\lab{dsubalg}
\ee
Note that in \rf{dsubalg} the differential operators are put to the
left. Henceforth, we shall skip the sign $\circ$ in symbol products for
brevity.

Defining $R_{\ell} = P_{+} - P_{-}$ for each of the three cases,
eqs.\rf{R-lie1}
take the form :
\be
\lb X , Y \rb_{R_{\ell}} = \lb X_{\geq \ell},
Y_{\geq \ell}\rb - \lb X_{< \ell} , Y_{< \ell} \rb \quad , \quad
ad^{\ast}_{R_\ell} (X) L = \left\lb X_{\geq \ell} , L_{< -\ell} \right\rb_{<
-\ell}
- \left\lb X_{< \ell} , L_{\geq -\ell} \right\rb_{\geq -\ell}   \lab{R-bra}
\ee

Now, choosing an infinite set of independent Casimir functions :
\be
H_{m+1} = { 1 \ov {m+1}} \Intres L^{m+1} \quad , \; m=0,1,2,\ldots
\lab{casimir}
\ee
the three decompositions \rf{dsubalg}
of $\PsDA\,$ labelled by $\ell=0,1,2$ yield, according to the AKS-RS scheme,
three different integrable hierarchies -- the standard KP hierarchy ($\ell =0$)
and the first and second {\em modified} KP hierarchies ($\ell =1,2$) :
\be
\partder{L}{t_m} + \Sbr{\( L^m \)_{\geq \ell}}{L} = 0    \lab{rkp}
\ee
One can show that all three KP hierarchies are related through symplectic gauge
transformations \ct{ANPW}. Here we shall concentrate on various {\em Poisson
reductions} \foot{Let us recall the general notions \ct{MR86}. Let
$\, \( \cM ,P\) \,$ be a smooth Poisson
manifold with Poisson structure $\, P \; : \; T^{\ast}(\cM ) \longrightarrow
T(\cM )\,$ and let $S$ be a smooth submanifold of $\cM\,$ with embedding
$\, \m \; :\; S \longrightarrow \cM $ . Now, a Poisson structure
$\, P^{\pr}\; : \; T^{\ast}(S) \longrightarrow T(S)\,$ on
$\, S \subset \cM\,$ is called {\em Poisson reduction} of
$\, P\,$ if for arbitrary
functions on $\cM\,$ the following property is satisfied :
$\,\m^{\ast} \( \lcurl f_1 ,\, f_2 \rcurl_P \) =
\lcurl \m^{\ast}f_1 ,\, \m^{\ast}f_2 \rcurl_{P^{\pr}} \,$.
In other words, restriction of the Poisson brackets w.r.t. $P$ of
arbitrary functions on $\cM\,$ to the submanifold $S\,$ is equivalent to
computing the Poisson brackets w.r.t. $P^{\pr}$ of the restrictions on
$S\,$ of these same functions.}
of the standard KP hierarchy which appear, in particular, in the
context of matrix models of strings.

The phase space of the standard KP integrable system is :
\be
\cM_{KP} = \lcurl L = D + \sum_{k=1}^{\infty} u_k (x) D^{-k} \rcurl
\lab{Lax-KP}
\ee
(it by itself is a trivial Poisson reduction from $\cM = \dPsDA \simeq \PsDA$).
The KK Poisson brackets (first eq.\rf{Rbra1}) read on \rf{Lax-KP} :
\br
\lcurl u_n(x)\, , \, u_m(y) \rcurl = \Omega_{n-1,m-1}(u(x)) \,
\d (x-y)   \phanta  \lab{Winf}  \\
\Omega_{nm}(u(x)) \equiv \sum_{k=0}^{n}(-1)^k
{n \choose k} u_{n+m-k+1}(x) D^k_x -
\sum_{k=0}^{m} {m \choose k} D^k_x
u_{n+m-k +1}(x)     \lab{omega}
\er
and are recognized as the (centerless) $\Win1\,$ algebra (see \ct{W-alg}),
which
is isomorphic to $\DA \subset \PsDA\,$ (algebra of purely differential
operators) \ct{DOP}.

Now, returning to the flow eqs. in (multi)matrix models \rf{23} with Lax
operator \rf{25}, one can show \ct{multikp} that the space of $2q$-boson
Lax operators :
\be
\cM_q = \lcurl L_q = D + \sum_{l=1}^q A_l \( D - B_l \)^{-1} \( D - B_{l+1}
\)^{-1} \ldots \( D - B_q \)^{-1}        \rcurl  \lab{2q-Lax}
\ee
is a legitimate {\em Poisson reduction} of the full KP hierarchy given by
\rf{Lax-KP}, {\sl i.e.}, the flow eqs.\rf{23} indeed correspond to Hamiltonian
integrable systems. Furthermore, what is even more important, one can express
all $2q$ fields parametrizing \rf{2q-Lax} in terms of {\em Darboux} canonical
pairs of coordinates $ \Bigl( a_r (x), b_r (x) \Bigr)_{r=1}^q$ :
\br
B_l = b_l + b_{l+1} + \cdots b_q  \qquad , \qquad
A_q = a_q  \phanta \lab{3-4}  \\
A_{q-r}(a,b) = \sum_{n_r =r}^{q-1} \cdot\cdot\cdot \sum_{n_2 =2}^{n_3 -1}
\sum_{n_1 =1}^{n_2 -1} \( \pa + b_{n_r} + \cdot\cdot\cdot +
b_{n_r -r +1} \) \cdot\cdot\cdot \( \pa + b_{n_1}\) a_{n_1} \lab{3-5}   \\
\lcurl a_r (x),\, b_s (y) \rcurl_{P^{\pr}} = - \d_{rs} \pa_x \d (x-y)
\quad \phanta \lab{3-2}
\er
Expressing $L_q$ in \rf{2q-Lax} as power series in $D^{-1}$ :
\br
L_q = D + \sum_{k=1}^{\infty}
U_k \lb (a,b) \rb (x) D^{-k}   \phantb  \lab{3-A}  \\
U_k \lb (a,b) \rb (x) = a_q P^{(1)}_{k-1}\( b_q\) +
\sum_{r=1}^{\min (q-1,k-1)} A_{q-r}(a,b) P^{(r+1)}_{k-1-r}
\bigl( b_q, b_q + b_{q-1}, \ldots ,
\sum_{l=q-r}^q b_l \bigr)   \lab{3-B}
\er
where $\,A_{q-r}(a,b)\,$ are the same as in \rf{3-5}, and
$\, P^{(N)}_n \,$ denote the (multiple) \faa polynomials :
\be
P^{(N)}_n (B_N ,B_{N-1},\ldots ,B_1) =
\sum_{m_1 + \cdot\cdot\cdot + m_N = n}
\( -\pa + B_1 \)^{m_1} \cdot\cdot\cdot \( -\pa + B_N\)^{m_N}
\cdot 1   \lab{multifaa}
\ee
one obtains a series of explicit (Poisson bracket) realizations of $\Win1$
algebra in terms of {\em finite} number of $2q$ bosons
$\, \( a_r ,b_r \)_{r=1}^q$ for any $q = 1,2,3,\ldots\,$
(cf. \rf{Winf}-\rf{omega}) :
\be
\Big\{ U_k \lb (a,b) \rb (x) \, ,\, U_l \lb (a,b) \rb (y) \Bigr\}
= \O_{k-1,l-1} \bigl( U \lb (a,b) \rb \bigr) \d (x-y) \lab{3-C}
\ee
Similar ``Darbouxization'' of $L_q$ \rf{2q-Lax} holds
w.r.t. the second KP Hamiltonian structure (see next subsection).
\mskp
{\bf 3.3 Lie-Poisson Groups and Lie Bi-Algebras : Hamiltonian approach}
\sskp
Completely integrable systems possess another notorious feature -- that of
{\em multi}-Hamil\-to\-ni\-an structure,
{\sl i.e.}, they always possess at least two
independent {\em compatible} Poisson structures \foot{Two Poisson structures
$\{ \cdot ,\cdot \}_1$ and $\{ \cdot ,\cdot \}_2$ are called {\em compatible}
if their linear combination $\{ \cdot ,\cdot \}_{\l} = \{ \cdot ,\cdot \}_1
+ \l \{ \cdot ,\cdot \}_2$ is also a Poisson structure.}.
In the case of KP
integrable hierarchies (and their reductions) it turns out that the ``second''
Poisson structure can be understood in terms of the {\em Lie-Poisson} structure
on the Lie group $G$ corresponding to the Lie algebra $\lie$ entering the
AKS-RS construction, whereas the ``first'' Poisson structure is just the KK
structure \rf{Rbra} on $\dlie_R$.

The notions of a {\em Lie-Poisson} group and its ``infinitesimal'' version --
Lie {\em Bi-algebra} were first introduced by Drinfeld \ct{L-P-drin}.
A Lie group $G$ is called {\em Lie-Poisson} if on the algebra of smooth
functions on it $Fun (G)$ there exists Poisson structure
which is compatible with the group multiplication, {\sl i.e.},
$\D \(\Pbr{F_1}{F_2}\) = \Pbr{\D \( F_1\)}{\D \( F_2\)}\,$ where $\D$ denotes
the coproduct in $Fun (G)\,$ $\bigl( \; \D F (g,h) = F(gh) \; \Bigr)$.
More precisely, the Lie-Poisson structure is given by :
\br
\Pbr{F_1 (g)}{F_2 (g)} =
\bil{\nabla_L F_1 (g) \otimes \nabla_L F_2 (g)}{r(g)} \phanta  \nonu    \\
\( \;\; = - \bil{\nabla_R F_1 (g) \otimes \nabla_R F_2 (g)}{r(g^{-1})}\;\)
\lab{LPbra}  \\
\bil{\nabla_L F(g)}{X} = {d \ov {dt}}\bgv_{t=0} F \( e^{tX}g\) \quad , \quad
\bil{\nabla_R F(g)}{X} = {d \ov {dt}}\bgv_{t=0} F \( g e^{tX}\)
\lab{Lie-der}
\er
where $\nabla_{L,R}$ denote left/right Lie-derivatives, and where $r(g)$ is a
{\em cocycle} on $G$ with values in $\lie \otimes \lie$ :
\be
r(gh) = r(g) + Ad(g)\otimes Ad(g) \, r(h)    \lab{r-cocycle}
\ee

In the case when $r(g)$ is a {\em coboundary} , {\sl i.e.}
\be
r(g) = Ad(g)\otimes Ad(g) \, r_0 - r_0    \lab{r-cobound}
\ee
with $r_0 \in \lie \otimes \lie$ being a constant element, one easily finds
that the Jacobi identity for \rf{LPbra}
reduces precisely to the YBE \rf{CYBE} for $r_0$ as a classical $r$-matrix.

Let us note that for matrix groups eq.\rf{LPbra} can be written in a simpler
form :
\be
\lcurl g \sto g \rcurl = r(g) \, g \otimes  g \qquad
\Bigl( \,\; \lcurl g \sto g \rcurl = - \sbr{r_0}{g \otimes  g}  \quad
{\rm in ~case ~of ~\rf{r-cobound}}  \Bigr)       \lab{LPbra1}
\ee

Eq.\rf{r-cocycle} implies the following exterior derivative equation :
\br
d\, r(g) = \llb \ot{Y(g)}\sta r(g) \rrb - \p \( Y(g)\)   \lab{dr} \\
\p (X) = - {d\ov {dt}}\bgv_{t=0} r \( e^{Xt} \) \quad {\rm for}
\quad \forall X \in \lie   \lab{phi}
\er
where $Y(g) = dg\, g^{-1}$ denotes the Maurer-Cartan one-form on $\lie\,$,
and $\p (\cdot ) $ defined in \rf{phi} is a $\lie \otimes \lie$-valued
cocycle on $\lie$ :
\be
\p \( \llb X \sta Y \rrb \) = \llb \ot{X} \sta \p (Y) \rrb +
\llb \p (X) \sta \ot{Y} \rrb     \lab{phi-cocycle}
\ee

The solution of \rf{dr} is a coboundary $r(g)$ \rf{r-cobound} if and only if
$\p (\cdot )$ is coboundary :
\be
\p (X) = \llb r_0 \sta \ot{X} \rrb    \lab{phi-cobound}
\ee
The cocycle $\p (\cdot )$ allows to introduce a Lie-commutator $\lb \cdot ,
\cdot \rb_{\ast}$ on the dual space $\dlie$ and, correspondingly, coadjoint
action $\dad (\cdot )$ of $\dlie$ on $\lie$ as follows :
\be
\bil{{\Sbr{U}{V}}_\ast}{X} \equiv - \bil{V}{\dad (U)X} =
\bil{U\otimes V}{\p (X)}
\qquad  \forall U,V \in \dlie \; ,\; \forall X \in \lie \lab{dual-bra}
\ee
whereas the mixed commutator between elements of $\lie$ and $\dlie$ is defined
as :
\be
\sbr{X}{U}= ad^{\ast} (X)U - \dad (U) X     \lab{mixed-bra}
\ee
According to an important theorem by Drinfeld \ct{L-P-drin}, the Lie algebra
$\lie$ of each Lie-Poisson group $G$ possesses a Lie-{\em bialgebraic}
structure
and {\em vice versa}.
Namely, by means of \rf{dual-bra} and \rf{mixed-bra} the direct sum (as vector
space)
$\cD = \lie \oplus \dlie$ becomes itself a Lie algebra called ``the double'',
such that $\lie$ and $\dlie$ are {\em isotropic} subalgebras of $\cD\,$ w.r.t.
the Killing form on $\cD$ :
\be
\Bigl( \( X_1 ,U_1 \)\, ,\, \( X_2 ,U_2 \) \Bigr) =
\me{U_1}{X_2} + \me{U_2}{X_1} \quad \forall \( X_{1,2} ,U_{1,2} \) \in \cD
\lab{Bil-double}
\ee
The triple $\Bigl( \cD ,\lie ,\dlie \Bigr)$ is also called {\em Manin triple}.
Furthermore, the
{\em double group} ${\wti D} \simeq G \times G^{\ast}$ , corresponding to $\cD$
,
is direct product (as a manifold) of its subgroups $G$ and $G^{\ast}$
corresponding
to $\lie$ and $\dlie$, respectively.
Let us stress the symmetry of the above construction under the exchanges $\lie
\longleftrightarrow \dlie \; , \; G \longleftrightarrow G^{\ast}$ .

Now it is easy to write down the explicit solution for the Lie-Poisson group
cocycle $r(g)\,$ \rf{r-cocycle} :
\be
\me{U \otimes V}{r(g)} =
\me{{\Bigl( g^{-1} V g \Bigr)}_{-}}{{\Bigl( g^{-1} U g \Bigr)}_{+}} \quad
\forall U,V \in \dlie   \lab{r-solution}
\ee
where the subscripts $(\pm)$ indicate projections in the double algebra $\cD$
along $\lie \, ,\, \dlie$ , respectively.

As in the AKS-RS Lie-algebraic scheme, one can construct integrable Hamiltonian
systems on Lie-Poisson groups. Namely, from \rf{r-cocycle} or \rf{r-cobound}
one
can easily verify that all $Ad (\cdot )$-invariant functions on $G$ ,
$H\lb h g h^{-1} \rb = H\lb g\rb\,$, are in involution w.r.t. the Lie-Poisson
structure \rf{LPbra} : $ \pbbr{H_k \lb g\rb}{H_l \lb g\rb} = 0$ . Similarly,
the analogues of the Hamiltonian eqs. of motion \rf{rhameqs} and the
associated geometric action \rf{R-action} take now the form :
\br
\partder{g}{t_k} g^{-1} = - {\hat r}_g \( \nabla_L H_k \lb g\rb \)
\phanta \lab{rhameqs-1}  \\
\cW \lb g\rb = -\h \int d^{-1} \( \bil{{\hat s}_g \bigl( Y(g)\bigr)}{Y(g)}\)
- \int dt \, H_k \lb g\rb   \lab{R-action-1}
\er
where the action of the operator ${\hat r}_g \, :
\, \dlie \longrightarrow \lie\,$ is defined by $\me{U}{{\hat r}_g \( V\)}
\equiv \me{U \otimes V}{r(g)}\,$, cf. \rf{r-solution},
$ {\hat s}_g = {\hat r}_g^{-1}$, and $d^{-1}$ denotes inverse operator of
the exterior derivative acting on the corresponding closed 2-form
\foot{Closeness follows from the exterior derivative equation $\; d{\hat s}_g
= ad^{\ast}\bigl( Y(g)\bigr) {\hat s}_g - {\hat s}_g ad\bigl( Y(g)\bigr) -
{\hat s}_g \p \bigl( Y(g)\bigr) {\hat s}_g\,$,
satisfied by the inverse cocycle operator ${\hat s}_g\,$ as a consequence of
eq.\rf{dr}.} (recall $Y(g) = dg g^{-1}\,$).

The principal example is provided by the {\em extended} Volterra algebra of
purely pseudo-differential operators ($c$ is arbitrary constant) :
\be
\lie \equiv \eVolt = \lcurl \sum_{k \geq 1} u_k (x) D^{-k} + c \ln D \rcurl
\lab{eVolt}
\ee
whose Lie-double is the {\em extended} algebra of all pseudo-differential
operators :
\be
\cD \equiv \ePsDA = \eVolt \oplus \eDA \quad ; \quad \dlie \equiv \eDA \simeq
\Win1 = \lcurl \sum_{l \geq 0} D^l v_l (x) + \a {\hat E} \rcurl    \lab{ePsDA}
\ee
Here ${\hat E}$ indicates the central element of $\Win1\,$, as well as of the
whole
$\ePsDA\,$, which is dual to $\ln D\,$ , cf. \ct{DOP,KhZa}. The corresponding
extended Volterra group (exponentiation of \rf{eVolt}) :
\be
G \equiv \eVOLT = \lcurl g \equiv L= \( 1 + \sum_{k \geq 1} {\ti u}_k (x)
D^{-k}\)
\circ D^c \rcurl        \lab{eVOLT}
\ee
can be viewed as a set of spaces (for each $c$ fixed)
of Lax operators of generalized KP hierarchies w.r.t.
the {\em second} KP Hamiltonian structure \ct{KP} (eq.\rf{second-KP} below).
The latter is given precisely by
the Lie-Poisson structure \rf{LPbra1} with the cocycle $r(g)\,$
\rf{r-solution} for $G = \eVOLT\,$ \rf{eVOLT} (see \ct{KhZa}).

Let us go back to the example of $2q$-boson KP Lax operators appearing in the
multi-matrix string models \rf{2q-Lax}. The second Hamiltonian structure
for general Lax operators \rf{Lax-KP} has the form\foot{The second term on the
r.h.s. of \rf{second-KP} is a Dirac bracket term due to the
{\em second class} constraint $u_0 =0$ in $L\,$ \rf{Lax-KP}.} :
\br
\pbbr{\me{L}{X}}{\me{L}{Y}} = {\Tr}_A \( \( LX\)_{+} LY -
\( XL\)_{+} YL \) +  \nonu  \\
\int dx \, {\rm Res}\Bigl( \sbr{L}{X}\Bigr) \pa^{-1}
{\rm Res}\Bigl( \sbr{L}{Y}\Bigr) \qquad     \lab{second-KP}
\er
Here ${\Tr}_A$ denotes the Adler trace \rf{adler} and the subscript $+$
indicates taking the purely differential part. For the coefficient fields
$u_k (x)$ of $L\,$ \rf{Lax-KP} the second KP Poisson algebra \rf{second-KP}
yields the {\em nonlinear} ({\sl i.e.}, non-Lie) $\nWinf$ algebra \ct{yu-wu}.
The latter appears as a unique (modulo certain homogenity assumptions)
nonlinear deformation of $\Win1$ algebra.

In analogy with eqs.\rf{3-4}--\rf{3-2} one can express \cite{recur}
the coefficient fields $\( A_l ,B_l \)_{l=1}^q$ of $L_q\,$ \rf{2q-Lax}
in terms of Darboux canonical pairs of fields
$\( c_r , e_r \)_{r=1}^q$ w.r.t. the second KP Hamiltonian structure
\rf{second-KP} :
\br
B_k = e_k +\sum_{l = k}^{q}  c_l \quad , \quad 1 \leq k \leq q \quad ; \quad
A_q =\sum_{r=1}^{q} \( \pa +c_r \) e_r \quad \lab{rec-1}  \\
A_k =  \sum_{n_k =1}^k \( \pa + e_{n_k} - e_{n_k +q-k}
+ \sum_{l_k = n_k}^{n_k +q-k} c_{l_k} \) \qquad \qquad \nonu  \\
\times  \sum_{n_{k-1}=1}^{n_k} \( \pa + e_{n_{k-1}} - e_{n_{k-1} +q-1-k}
+ \sum_{l_{k-1} = n_{k-1}}^{n_{k-1} +q-1-k} c_{l_{k-1}} \) \times \cdots
\qquad \nonu \\
\times \sum_{n_2 =1}^{n_3} \( \pa + e_{n_2} - e_{n_2 +1} +
c_{n_2} + c_{n_2 +1} \) \, \sum_{n_1 =1}^{n_2} \( \pa + c_{n_1} \) e_{n_1}
\quad , \; 1 \leq k \leq q-1   \lab{rec-2} \\
\lcurl c_k(x) \, ,\, e_l(y) \rcurl = - \d_{kl} \pa_x \d (x-y) \quad , \;
k,l =1,2, \ldots, q  \qquad \lab{rec-3}
\er
These equations are equivalent to the following ``dressing'' form
for the $2q$-boson KP Lax operator :
\br
L_q = D + \sum_{l=1}^q A_l \( D - B_l \)^{-1} \( D - B_{l+1} \)^{-1} \ldots
\( D - B_q \)^{-1} =
\cU_q \ldots \cU_1 \, D\, \cV_1^{-1} \ldots \cV_q^{-1}  \lab{rec-4}  \\
\cU_k \equiv \( D - e_k \) e^{\int c_k} \quad , \quad
\cV_k \equiv e^{\int c_k} \( D - e_k \)  \quad , \; \, k=1,\ldots ,q
\quad \lab{rec-5}
\er
Eqs.\rf{rec-1}--\rf{rec-3} or, equivalently, eqs.\rf{rec-4}--\rf{rec-5} can
be viewed as {\em generalized Miura transformation} for the $2q$-boson KP
hierarchy \foot{For discussion of the generalized Miura transformation and
the associated Kuperschmidt-Wilson theorem, we refer to \ct{KP}.}.

The Miura-transformed form of $L_q$ \rf{rec-4} reads explicitly :
\br
L_q = D + \sum_{k=1}^{\infty}
U_k \lb (c,e) \rb (x) D^{-k}   \phantb  \lab{rec-A}  \\
U_k \lb (c,e) \rb (x) = P^{(1)}_{k-1} \( e_q + c_q \)
\sum_{l=1}^q \( \pa + c_l \) e_l + \qquad\quad  \nonu \\
\sum_{r=1}^{\min (q-1,k-1)} A_{q-r}(c,e) P^{(r+1)}_{k-1-r}
\bigl( e_q + c_q , e_{q-1} + c_{q-1} + c_q , \ldots ,
e_{q-r} + \sum_{l=q-r}^q c_l \bigr)  \lab{rec-B}
\er
where $\,A_{q-r}(c,e)\,$ are the same as in \rf{rec-2}, and
$\, P^{(N)}_n \,$ denote the (multiple) \faa polynomials \rf{multifaa}.

Now, in complete analogy with eqs.\rf{3-A}--\rf{3-C}, which yield a
series of realizations of the linear $\Win1$ algebra in terms of $2q$ bosons,
we obtain, upon substitution of \rf{rec-A}--\rf{rec-B} into \rf{second-KP},
a series of explicit (Poisson bracket)
realizations of the nonlinear $\nWinf$ algebra in terms of $2q$ bosonic fields
for any $q=1,2, \ldots$ . This algebra plays an important r{\^{o}}le
as a ``hidden'' symmetry algebra in string-theory-inspired
models with black hole solutions \ct{yu-wu}.

Concluding this section, let us note that in the general case Lie-Poisson
groups provide natural
geometric description of the {\em dressing symmetries} in completely integrable
models \ct{dress,Alekseev}.
Another outstanding r{\^{o}}le played by Lie-Poisson
structures \rf{LPbra1} is their appearance in the context of the classical
inverse scattering method \ct{FT87} as fundamental {\em Sklyanin brackets}
for the monodromy matrix $g \simeq T(\l )\,$ of the auxiliary linear spectral
problem.
\lskip
{\large {\bf 4. Quantum Integrable Models}}
\mskp
{\bf 4.1 Quantization of Lie-Poisson Groups : Quantum Groups}
\sskp
Historically, quantization of completely integrable models, whose Hamiltonian
structure is based on group Lie-Poisson structures, laid for the first time
to explicit construction, in the context of the quantum version of the
inverse scattering method \ct{QISM}, of {\em quantum groups} which were
subsequently identified with {\em quasi-triangular Hopf algebras} \ct{QG}.

Among the various ways to introduce quantum groups there exists an approach
\ct{Takht90}, whose conceptual point of view underscores both the {\em quantum
mechanical} as well as the {\em Hopf algebraic} aspects in quantization of
Lie-Poisson groups. Namely, on one hand $Fun (G)$ can be viewed as Abelian
associative algebra of ``observables'' of a classical Hamiltonian system
$( \cM ,\cP )$ with a phase space $\cM = G$ and Poisson structure $\cP =
\cP_{LP}\,$ given by \rf{LPbra} :
\be
\cP_{LP} \( F_1 ,F_2\) \equiv \pbbr{F_1}{F_2} =
\me{\nabla_L F_1 \otimes \nabla_L F_2 - \nabla_R F_1 \otimes \nabla_R F_2}{r_0}
\lab{LP-cobound}
\ee
where the classical $r$-matrix $r_0$ satisfies the classical YBE \rf{CYBE}
(from now on we shall consider only coboundary Lie-Poisson groups
\rf{r-cobound}). On the other hand, one can easily check that $Fun (G)$ is
endowed with a structure of a {\em commutative}, but {\em non}-cocommutative,
Hopf algebra $\cA_0 (m,\D ,S,\vareps ) \equiv Fun (G)$ with a product
$m \( F_1 ,F_2 \) (g) = F_1 (g) F_2 (g)$ ,
coproduct $ \D \bigl( F\bigr) (g_1, g_2 ) = F (g_1 g_2)$ ,
antipode $ SF (g) = F(g^{-1})$ and counit $ \vareps (F) = F(e)$ ,
and this Hopf structure is compatible with the Poisson structure
\rf{LP-cobound}, {\sl i.e.}, $\D \circ \cP = \cP \circ \D$ .

Thus, quantization of a Lie-Poisson group $G$ may be viewed as a
generalization of Weyl quantization $Fun (G) \longrightarrow Fun_h (G)$
of a classical Hamiltonian system defined by $\(\cM \equiv G , \cP_{LP} \)$ ,
{\sl i.e.}, {\em non-commutative} deformation
of the product $m (\cdot ,\cdot) \longrightarrow m_h (\cdot ,\cdot ) $ with
a deformation parameter $h$, which satisfies the additional condition that
the deformed algebra $Fun_h (G) \equiv \cA_h (m_h ,\D ,S,\vareps )$ is
again (non-commutative and non-cocommutative) Hopf algebra which is a
deformation of $\cA_0 (m,\D ,S,\vareps ) \equiv Fun (G)$ .

Let us recall \ct{Flato}, that in ordinary Hamiltonian mechanics on a Poisson
manifold $(\cM ,\cP )\,$ with local coordinates $\( x^i \)$ and constant
Poisson tensor $\pbbr{x^i}{x^j} = P^{ij}$ , Weyl quantization is given by the
associative and non-commutative {\em Moyal} product :
\br
m_h \( F_1 ,F_2 \) \equiv F_1 \star_h F_2 = m \circ e^{{h\ov 2}\cP}
\( F_1 ,F_2\) = F_1 \cdot F_2 + {h\ov 2} \pbbr{F_1}{F_2} + O(h^2 )
\lab{Moyal}  \\
\cP = P^{ij} \partder{}{x^i} \otimes \partder{}{x^j} \; : \;
Fun(\cM ) \otimes Fun(\cM ) \longrightarrow Fun(\cM ) \otimes Fun(\cM )
\lab{bi-vector}
\er
where $\cP$ is the Poisson bi-vector field. Recall also, that the form of
the first order term in the $h$-expansion (last eq.\rf{Moyal}) is dictated by
the {\em semiclassical correspondence} principle.

In the case of Lie-Poisson groups $\( G, \cP_{LP} \)$ ,
the deformed product $m_h (\cdot ,\cdot )$
preserving the Hopf algebra structure and satisfying the semiclassical
condition, can be constructed as follows \ct{Takht90}.
Let us choose a basis
$\lcurl X^i \rcurl$ in $\ulie$ -- the universal enveloping algebra of
the Lie algebra $\lie$ of $G$,
and let $\pi_{L,R}\,$ denote the representations of
$\ulie$ in terms of left/right Lie derivatives : $\pi_{L,R} \( X^i \) =
\nabla^i_{L,R}\,$ (see eq.\rf{Lie-der}). Then :
\br
m_h (\cdot ,\cdot ) = m \circ {\wti \L} \quad , \quad
{\wti \L} = \( \pi_L \otimes \pi_L \) \bigl( \L \bigr) \circ
\( \pi_R \otimes \pi_R \) \bigl( \L^{-1} \bigr)   \lab{Moyal-LP} \\
\L (X,Y) = \sum_{\{\a\},\{\b\}} c_{\{\a\},\{\b\}}(h) \prod_{i=1}^{dim\,\lie}
\( X^i \)^{\a_i} \prod_{j=1}^{dim\,\lie} \( Y^j \)^{\b_j} = \one +
{h\ov 2} r_{ij} X^i Y^j + O(h^2 )    \lab{Lambda}
\er
with the following notations. The coefficients in $\L (\cdot ,\cdot) \; : \;
\ulie \otimes \ulie \longrightarrow \ulie \otimes \ulie \lb\lb h\rb\rb\,$
are power series in
$h$ ; $\{ X^i \}$ and $\{ Y^j \}$ are generator basises in the first and
second copy of
$\ulie$, respectively; $\| r_{ij}\| = r_0$ is just the classical $r$-matrix
satisfying \rf{CYBE}. Moreover, the associativity condition for
$ m_h (\cdot ,\cdot ) \,$ \rf{Moyal-LP} implies the following basic quadratic
equation on $\ulie \otimes \ulie \otimes \ulie \lb\lb h\rb\rb\,$
( $X,Y,Z$ below correspond to
the first, second and third factor $\ulie$ in the tensor product) :
\be
\L (X+Y,Z) \L (X,Y) = \L (X,Y+Z) \L (Y,Z) \quad ; \quad
\L (X,0) = \L (0,Y) = \one   \lab{Lambda-eq}
\ee
Defining $ {\bar R}(X,Y) = \L^{-1} (Y,X) \L(X,Y) \in
\ulie \otimes \ulie \lb\lb h\rb\rb\,$, one obtains from \rf{Lambda-eq} :
\br
{\bar R}(X,Y) {\bar R}(X,Z) {\bar R}(Y,Z) =
{\bar R}(Y,Z) {\bar R}(X,Z) {\bar R}(X,Y)     \lab{QYBE-univ}  \\
{\bar R}(X,Y) {\bar R}(Y,X) = \one  \quad ; \quad
{\bar R}(X,Y) = \one + h\; r_{ij} X^i Y^j + O(h^2 )   \nonu
\er
${\bar R}$ is called {\em universal quantum R-matrix} associated with the
classical $r$-matrix $\| r_{ij}\| = r_0$ . If $\rho \; :\; \lie \longrightarrow
End (\cV )$ is some representation of $\lie$ in a (finite-dimensional)
vector space $\cV\,$, then the matrix $R = (\rho \otimes \rho )\({\bar R}\)
\in End (\cV \otimes \cV )\,$ satisfies the famous {\em quantum Yang-Baxter}
equation (QYBE) (plus the ``unitarity'' condition,
$P \in End (\cV \otimes \cV )\,$ being the permutation operator) :
\be
\stackrel{(12)}{R} \stackrel{(13)}{R} \stackrel{(23)}{R} =
\stackrel{(23)}{R} \stackrel{(13)}{R} \stackrel{(12)}{R}  \qquad ; \quad
RPRP = \one   \quad ; \quad  R = \one + h\, r_0 + O(h^2 )  \lab{QYBE} \\
\ee
The indices $(12),(13),(23)$ indicate the various embeddings of
$R \in End (\cV \otimes \cV )\,$ in $End (\cV \otimes \cV \otimes \cV)\,$.

In particular, from the deformed product \rf{Moyal-LP} $m_h \(F_1 ,F_2 \) =
m \circ {\wti \L} \(F_1 ,F_2 \)\,$  where the functions $F_1 (g) = g_{ab} \,
,\, F_2 (g) = g_{cd}\,$ are just matrix elements of the group element $g \in G$
(for matrix groups), one obtains using matrix tensor notations :
\be
R (g \otimes \one ) (\one \otimes g) = (\one \otimes g) (g \otimes \one ) R
\lab{fund-CR}
\ee
The semiclassical limit of \rf{fund-CR} is precisely given by the Lie-Poisson
bracket \rf{LPbra1}. Eq.\rf{fund-CR} is nothing but the famous fundamental
commutation relations for the matrix elements of the quantum monodromy
matrix (with spectral parameter dependence suppressed) in the quantum inverse
scattering method \ct{QISM}.

For various treatments and numerous applications of QYBE, see ref.\ct{Jimbo}.
For parallel developments in the abstract Hopf algebraic context, we refer to
\ct{Mia}.
\mskp
{\bf 4.2 ``Soliton'' Scattering in Completely Integrable Models}
\sskp
Let us now consider integrable $D=2$ relativistic field theories whose
Hamiltonian dynamics
is given by actions $ S\lb \p \rb = \int d^2 x \, \cL (\p ,\pa \p )\,$ --
local functionals of the fundamental fields (collectively denoted by
$\p (x^{+},x^{-})\,$) and their derivatives. As usual, one uses the light-cone
form of the space-time coordinates : $x^{\pm} = \h \( x^1 \pm x^0 \)$ .
Complete integrability in this context implies the existence of an infinite
number of independent integrals of motion in involution $Q^{(s)}$ ,
whose densities are
local (as functionals of $\p$ and its derivatives) conserved currents :
\be
Q^{(s)} = \oint \( T^{(s+1)} dx^{-}
+ \Th^{(s-1)} dx^{+} \) \;\; , \; s=1,2,\ldots \quad ; \quad
\pa_{+} T^{(s+1)} = \pa_{-} \Th^{(s-1)}  \lab{conserv}
\ee
(here $\, s\,$ indicates the $D=2$ Lorentz weight).
Thus, quantization of completely integrable field theories means fulfilment
of the quantum {\em renormalized} Ward identities for the {\em renormalized}
quantum conserved currents $\( {\wti T}^{(s+1)},{\wti \Th}^{(s-1)} \)$ :
\be
\pa_{+} \llangle {\wti T}^{(s+1)}(x) \p (x_1 ) \ldots \p (x_n ) \rrangle -
\pa_{-} \llangle {\wti \Th}^{(s-1)}(x) \p (x_1 ) \ldots \p (x_n ) \rrangle =
\d-{\rm function ~terms} \lab{ward}
\ee
(as usual, $\langle \ldots \rangle\,$ here denote time-ordered
correlation functions).
The infinite set of Ward identities lead to severe restrictions on the
particle (``soliton'') scattering processes -- conservation of all (odd)
powers of momenta of incoming and outgoing particles :
\be
\sum_{l=1}^{N_{in}} p_{l(in)}^{2n+1} = \sum_{l=1}^{N_{out}} p_{l(out)}^{2n+1}
\lab{momenta}
\ee
which, in turn, implies \ct{currents} : (a) {\em no} multi-particle production,
{\sl i.e.}, $N_{in} = N_{out}$ ; and (b) {\em factorization} of multi-particle
scattering amplitudes. The latter property is of tremendous importance, as it
leads to the remarkable Zamolodchikov's factorization eqs. for the $3$-particle
amplitudes \ct{Zam's}, meaning that any $3$-particle scattering process is
accomplished as a sequence of $2$-particle scatterings only and, moreover,
the amplitude does not depend on the order in which these sequential
$2$-particle scatterings occur:
\be
S_{i_1 i_2}^{k_1 k_2} \( \th_{12}\) \, S_{k_1 i_3}^{j_1 k_3} \( \th_{13}\) \,
S_{k_2 k_3}^{j_2 j_3} \( \th_{23}\)
= S_{i_2 i_3}^{k_2 k_3} \( \th_{23}\) \, S_{i_1 k_3}^{k_1 j_3} \( \th_{13}\)
\, S_{k_1 k_2}^{j_1 j_2} \( \th_{12}\)   \lab{factoriz}
\ee
with $\, \th_{ab} \equiv \th_a - \th_b \; ,\; a,b = 1,2,3$ .
In \rf{factoriz} the following notations are used : $S_{ij}^{kl} \(
\th_{12}\)\,$
denotes $2$-particle scattering amplitude of incoming particles of ``type"
labelled by the indices $i$ and $j$ and with (on-mass-shell) momenta
$p_{1,2} = m_{1,2} \( \cosh \th_{1,2},\sinh \th_{1,2} \)$, respectively
( $\th_{1,2}$ are the relativistic ``rapidities'').

Now, denoting by $\cV\,$ the vector space of internal particle symmetry
(particle ``types''), one can regard the matrix of the $2$-particle amplitude
as :
\br
S \(\th_{12}\) = \| S_{ij}^{kl} \( \th_{12}\) \| \in
Mat (\cV ) \otimes Mat (\cV )  \quad ({\rm for ~fixed} \; \th_{12})
\lab{S=R}  \\
\stackrel{(12)}{S} \(\th_{12}\) \equiv S \(\th_{12}\) \otimes \one \in
Mat (\cV ) \otimes Mat (\cV ) \otimes Mat (\cV )   \nonu
\er
and, accordingly, for $ \stackrel{(13)}{S} \(\th_{13}\)\,$ and
$\stackrel{(23)}{S} \(\th_{23}\)\,$ . Then it is straightforward to identify
\rf{factoriz} with the QYBE \rf{QYBE} in the quantum group framework.

A closely related natural appearance of quantum group structure in ``soliton''
scattering is provided by the notion of asymptotic states' symmetry
\ct{QG-soliton} : $ | (\th ,i)\rangle \longrightarrow T_{ij}(\th )
| (\th ,j)\rangle \;\, ;\; | (\th ,i)\rangle $
$\in \cV \; ,\; \| T_{ij}(\th ) \| \in Mat (\cV ) $ ,
as a result of the integrability. Namely, the $2$-particle $S$-matrix :
\be
| \left. \( \th_1 ,i_1\)\, ,\,\( \th_2 ,i_2\) \rrangle^{in} =
S_{i_1 i_2}^{j_1 j_2} \( \th_{12}\)
| \left. \( \th_2 ,j_2\)\, ,\,\( \th_1 ,j_1\) \rrangle^{out}
\lab{asympt}
\ee
can be viewed as a mapping (for fixed ``rapidities'')
$\,\stackrel{(12)}{S} \, : \; \stackrel{(1)}{\cV} \otimes \stackrel{(2)}{\cV}
\longrightarrow \stackrel{(2)}{\cV} \otimes \stackrel{(1)}{\cV}\;$
and, therefore, its invariance under the asymptotic states's symmetry :
\be
\stackrel{(12)}{S} \( \th_{12}\) \stackrel{(1)}{T} \( \th_{1}\)
\stackrel{(2)}{T} \( \th_{2}\) = \stackrel{(2)}{T} \( \th_{2}\)
\stackrel{(1)}{T} \( \th_{1}\) \stackrel{(12)}{S} \( \th_{12}\) \lab{QG-sol}
\ee
is straightforwardly identified with the structural relations \rf{fund-CR} for
quantum groups.

Let us point out that quantum group relations of exactly the same form as
\rf{factoriz} and \rf{QG-sol} do appear in exactly solvable lattice models of
planar statistical mechanics, but in this case -- with purely imaginary
``rapidities'' ( $ \th = i \a$ , $\a$ being
angles characterizing the rectangular lattices), $S_{ij}^{kl} \( \a\)\,$
being the matrix of Boltzmann weights at each lattice vertex, and
$T_{ij}(\a )\,$ denoting the row transfer matrix \ct{Zam80}.
\mskp
{\bf 4.3 Quantum Field Theory Approach to Integrable Models with
Dynamically Broken Conformal Invariance}
\sskp
Finally, let us briefly discuss construction of higher local quantum conserved
currents fulfilling the Ward identities \rf{ward}, which is the heart of the
quantum field theory approach to quantization of completely integrable models.
Recently, Zamolodchikov \ct{Zam89} proposed powerful general formalism based on
treating integrable models as {\em mass perturbations} of conformal field
theories : $S \lb \p \rb = S_{conf} \lb \p \rb + \sum_i m_i \int d^2 x \,
B_i ( \p ,\pa \p )\,$, where the ``coupling'' constants $m_i$ have positive
mass dimensions and $B_i ( \p ,\pa \p )\,$ are composite fields with conformal
dimensions less than 2. Since in general it is not possible to find explicit
expressions for $S_{conf} \lb \p \rb\,$ and $ B_i ( \p ,\pa \p )\,$ as
local functionals of local fundamental fields $\{ \p \}$, Zamolodchikov's
approach is purely algebraic using results from representation theory of
Virasoro algebra, in particular, information about the spectrum of conformal
field dimensions.

There exist, however, interesting classes of $D=2$ integrable field theories,
{\sl i.e.}, $O(N)$ nonlinear sigma-models and their supersymmetric
generalizations
with Lagrangians :
\br
\cL_{NL\s} = \h \pa_{+} n^a \pa_{-} n_a \quad , \quad {\vec{n}}^2 = N/g
\quad \; ; \quad \vec{n} = \( n^1 ,\ldots n^N \)  \lab{NLsigma}  \\
\cL_{susy-NL\s} = \h \pa_{+} n^a \pa_{-} n_a +
i {\bar \psi}^a \g^\m \pa_\m \psi_a
- {g\ov N}\( {\bar \psi}^a \psi_a \)^2 \quad , \quad {\vec{n}}^2 = N/g \quad ,
\quad n^a \psi_a =0  \lab{susy-NLsigma}
\er
which are conformally invariant on the classical level but, upon quantization,
they undergo dynamical {\em dimensional transmutation}, manifested through
dynamical {\em mass generation}, leading to anomalous conformal symmetry
breakdown. This clearly precludes the use of conformal perturbation approach
to \rf{NLsigma}, \rf{susy-NLsigma}. Fortunately, there exist long ago
alternative
nonperturbative \foot{The term ``nonperturbative'' refers to expansions
different from (or, {\sl e.g.}, partial resummations of) the
ordinary perturbation theory w.r.t. the coupling constant $g$ in
\rf{NLsigma} and \rf{susy-NLsigma} which is plagued by infrared
divergences in $D=2$.} treatment of quantum field theory models with $O(N)$
or $SU(N)$ internal symmetry -- the $1/N$ expansion \ct{1/N}.

Let us briefly illustrate the construction \ct{AKNP} of higher local quantum
conserved currents for \rf{NLsigma} \ct{Pol77} within the $1/N$ expansion
framework (the same techniques applies to other $1/N$-expandable integrable
models as well). The $1/N$ expansion is obtained from the generating
functional of time-ordered correlation functions :
\br
Z \lb J\rb = \int \cD\vec{n}\, \prod_x \d \( {\vec{n}}^2 - N/g \) \,
\exp \lcurl i\int d^2 x \, \llb \h (\pa \vec{n})^2 + \(\vec{J},\vec{n}\)
\rrb \, \rcurl =  \nonu \\
\int \cD \s \,\exp \lcurl -{N \ov 2} S_1 \lb \s\rb +
{i\ov 2} \int d^2 x\, d^2 y\, \( \vec{J}(x), (-\pa^2 + \s )^{-1} \vec{J}(y) \)
\,\rcurl  \lab{n-action}  \\
S_1 \lb \s\rb \equiv \Tr \ln (-\pa^2 + \s ) + {i\ov g} \int d^2 x \, \s
\phanta \lab{eff-action}
\er
by expanding the effective $\s$-field action \rf{eff-action} around its
stationary point $\s_c \equiv m^2 = \m^2 e^{-4\pi/g}$
(dynamically generated mass of the ``Goldstone'' field $\vec{n}$ , $\m$ being
the renormalization scale), {\sl i.e.}, $\s (x) = m^2 +
{1\ov{\sqrt{N}}} {\ti \s}(x) $ . As
a result, one arrives at the $1/N$ diagram technique with (free) propagators
in momentum space :
\br
\llangle n^a \, n^b \rrangle_{(0)} = -i \( m^2 + p^2 \)^{-1} \d^{ab}
\quad , \quad
\llangle {\ti \s} \, {\ti \s} \rrangle_{(0)} = \Bigl( \S \(p^2 \) \Bigr)^{-1}
\lab{1/N-diag}   \\
\S \(p^2 \) = \int { {d^2 k} \ov {(2\pi )^2}} \,
\frac{1}{\( m^2 + k^2 \) \( m^2 + (p-k)^2 \)}   \qquad  \nonu
\er
and ordinary tri-linear ${\ti \s}nn$-vertices.

It turns out \ct{ArefNP}, that the $1/N$
expansion can be renormalized by adapting the well-known {\em BPHZ}
\foot{Bogoliubov-Parasiuk-Hepp-Zimmermann \ct{BPHZ}. As shown in
\ct{ArefNP}, the (supersymmetric) nonlinear sigma-models \rf{NLsigma} and
\rf{susy-NLsigma} are renormalizable within the $1/N$ expansion also in $D=3$
space-time dimensions in spite of their naive nonrenormalizability w.r.t. the
ordinary coupling constant perturbation theory (note e.g. the presence of the
four-fermion term in \rf{susy-NLsigma}).} renormalization technique. Another
remarkable property of the $1/N$ expansion for \rf{NLsigma} is that, in spite
of the manifest {\em linear} $O(N)$ symmetry of \rf{1/N-diag}, the nonlinearity
of the ``Goldstone'' field $\vec{n}(x)$ is preserved on the quantum level as an
identity on the correlation functions :
\be
\llangle \cN \Bigl\lb {\vec{n}}^2 P(\vec{n}, \pa \vec{n}) \Bigr\rb (x)
\ldots \rrangle =
{\rm const} \llangle \cN \Bigl\lb P(\vec{n}, \pa \vec{n}) \Bigr\rb (x)
\ldots \rrangle                   \lab{chiral}
\ee
where $P(\vec{n}, \pa \vec{n})\,$ is arbitrary local polynomial of the
fundamental fields
and their derivatives, and $\cN \lb \ldots \rb\,$ indicates renormalized
normal product of the corresponding composite fields.

Armed with the above machinery, the first higher quantum conserved
current (for $s=3\,$ in the notations of \rf{conserv}, \rf{ward})
takes the following form :
\be
{\ti T}^{(4)} = \cN \llb\( \pa_{-}^2 \vec{n} \)^2 \rrb +
a_1 \cN \llb\( (\pa_{-} \vec{n} )^2 \)^2 \rrb \quad , \quad
{\ti \Th}^{(2)} = \( \h + a_2 \) \cN \llb\(\pa_{-} \vec{n} \)^2 \s\rrb +
a_3 \pa_{-}^2 \s   \lab{3-current}
\ee
where all coefficients $a_{1,2,3} = O (1/N)\,$ are expressed in terms of
one-particle irreducible correlation functions and their derivatives
in momentum space at zero external momenta. Their explicit form can be
found order by order in $1/N$ from the renormalized $1/N$-diagram
technique described above \ct{AKNP}.

Let us stress, that the higher quantum conserved currents \rf{3-current} and
those for $s=5,7,\ldots\,$ do not have
analogues in the classical conformally invariant theory \ct{Pohl76}.

\end{document}